\documentclass[twocolumn,aps,prl,showpacs]{revtex4} 
\usepackage{graphicx}
\usepackage{color}
\usepackage{dcolumn}
\usepackage{bm}
\newcommand\beq{\begin{equation}}
\newcommand\eeq{\end{equation}}
\newcommand\bear{\begin{eqnarray}}
\newcommand\eear{\end{eqnarray}}
\begin{document}
\title{Isostaticity of Constraints in Jammed Systems of Soft Frictionless Platonic Solids}
\author{Kyle C. Smith}
\author{Timothy S. Fisher}
\email{tsfisher@purdue.edu}
\affiliation{Birck Nanotechnology Center and School of Mechanical Engineering,\\Purdue University, West Lafayette, Indiana 47906, USA}
\author{Meheboob Alam}
\affiliation{Engineering Mechanics Unit,\\Jawaharlal Nehru Centre for Advanced Scientific Research, Jakkur P.O., Bangalore 560064, India}
\begin{abstract}
{The average number of constraints per particle $\langle C_{total} \rangle$ in mechanically stable systems of Platonic solids (except cubes) approaches the isostatic limit at the jamming point ($\langle C_{total} \rangle \rightarrow 12$), though average number of contacts are hypostatic.  By introducing angular alignment metrics to classify the degree of constraint imposed by each contact, constraints are shown to arise as a direct result of local orientational order reflected in edge-face and face-face alignment angle distributions.  With approximately one face-face contact per particle at jamming chain-like face-face clusters with finite extent form in these systems.}
\end{abstract}
\pacs{45.70.Cc, 45.50.-j, 61.43.-j}
\maketitle



Following Maxwell's approach \cite{MaxPhiMag1864}, jammed assemblies of frictionless spheres exhibit average number of contacts per particle $\langle Z_{total} \rangle$ equal to the isostatic value $2n_{f}$ \cite{MoukPRL1998}, where $n_{f}$ is the degrees of freedom per particle. In contrast, ellipses \cite{MailPRL2009,DonPRE2007}, ellipsoids \cite{DonPRE2007}, tetrahedra \cite{JaoPRL2010,SmiAlaPRE2010}, and the remaining Platonic solids \cite{SmiAlaPRE2010} exhibit hypostatic behavior ($\langle Z_{total} \rangle<2n_{f}$).  The isostatic condition has been linked to the mechanical stability of soft sphere systems \cite{HernPRE2003}, and the hypostaticity of ellipses has been attributed to the presence of floppy vibrational modes, which provide vanishing restoring force \cite{MailPRL2009}.  Jaoshvili \textit{et al.} \cite{JaoPRL2010} recently asserted that the average constraint number $\langle C_{total} \rangle$, which incorporates topology-dependent contact constraint (\textsl{e.g.}, through vertex-face, edge-edge, edge-face, and face-face contact topologies), is isostatic for tetrahedral dice even though $\langle Z_{total} \rangle<2n_{f}$.  Similar approaches incorporating variable contact constraint have been utilized in the prediction of mechanism mobility as early as 1890 (see \cite{GogMMT2005}).  The presupposition underlying constraint counting as a valid means of predicting net degrees of freedom is the mechanical independence of constraints.  As a result, such constraint counting approaches fail to predict the true mobility of mechanisms when kinematic redundancy is present (see \cite{GogMMT2005}), and isostatic $\langle C_{total} \rangle$ is not guarranteed at the jamming point.  Though constraint counting may yield an isostatic result, the validity of this conclusion depends on the methods used to estimate $\langle C_{total} \rangle$ as we show in this Letter.  

The main goal of this Letter is to assess the validity of generalized isostaticity \cite{JaoPRL2010} for jammed systems of Platonic solids (except cubes) through energy-based structural optimization and objective topological classification.  We also present jamming threshold density and show that these packings are \textit{vibrationally stable}.  The ill-conditioned vibrational spectrum is a direct result of topological heterogeneity in the contact network, and we therefore identify and highlight the orientational order of contacts by introducing angular alignment metrics.  Topology classification via angular alignment metric distributions is thereby used to assess the \textit{isostaticity of constraints}.


\textit{Contact model and jamming protocol.} Structural optimization coupled with controlled consolidative and expansive strain is used to probe the jamming point as in \cite{SmiAlaPRE2010}.  The conservative model employed assumes that contact between particles $\alpha$ and $\beta$ results in energy $E_{\alpha\beta} = YV^{2}/4V_{p}$ after a Hookian contact model applied to uniaxially compressed bars, where $V$ is the intersection volume between the particles, $V_{p}$ is the volume of a single particle, and $Y$ is the elastic modulus.  Conjugate gradient minimization with line searching is utilized with a relative energy change convergence tolerance less than $10^{-12}$ at each strain step to simulate static equilibrium (see \cite{SmiAlaPRE2010}).  Density is defined as $\phi = NV_{p}/V_{cell}$, where $V_{cell}$ and $N$ are the volume and the number of particles in the primary periodic cell.


\begin{table}[ht] 
\caption{Jamming threshold $\phi_{J}$ estimated for each system with 95 \% confidence intervals.}
\scriptsize
\begin{tabular}{cccc}
\hline
\hline
  & $N = 100$ & $N = 400$ & exp. \cite{BakKudPRE2010}\footnotemark[1] \\
\hline
  Tetrahedra & 0.629$\pm$0.001 & 0.634$\pm$0.011 & 0.64$\pm$0.01\\
  Octahedra & 0.6796$\pm$0.0003 & 0.686$\pm$0.001 & 0.64$\pm$0.01\\
  Icosahedra & 0.6953$\pm$0.0003 & 0.7008$\pm$0.0003 & 0.59$\pm$0.01\\
  Dodecahedra & 0.7065$\pm$0.0002 & 0.7085$\pm$0.0003 & 0.63$\pm$0.01\\ 
\hline
\hline
\end{tabular}
\footnotetext[1]{Values presented in \cite{BakKudPRE2010} for tetrahedral dice.}
\end{table}

\begin{figure}
\includegraphics[width=4.1cm]{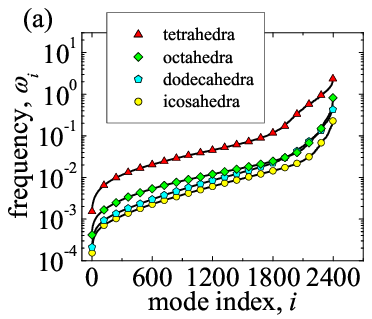} 
\includegraphics[width=4.1cm]{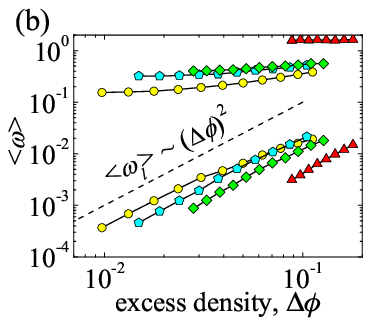} 
\caption{(Color online) (a) Vibrational spectra for stable systems nearest the jamming threshold.  (b) The power law scaling of $\langle \omega_{l} \rangle$ (lower curve set) and the asymptotically constant scaling of $\langle \omega_{h} \rangle$ (upper curve set).}
\end{figure}

Assemblies of monodisperse Platonic solids with periodic cubic lattice boundary conditions were consolidated with average energy per particle near $3.2 \times 10^{-5} YV_{p}$ from low-density random configurations at $\phi = 0.05$.  Estimates of the jamming threshold density $\phi_{J}$ (Table I) were obtained by expansion toward the jamming point as in \cite{SmiAlaPRE2010}.  $\phi_{J}$ converges well at $N=100$, but the results differ somewhat from the `random close packed' densities measured by \cite{BakKudPRE2010} for finite systems of rounded dice.  Unless otherwise stated, hereafter the results presented are for $N=400$.


\textit{Mechanical stability.} We computed the low-energy vibrational spectra of these monodisperse systems as $\omega_{i} = \sqrt{\lambda_{i}/m}$ (as in \cite{MailPRL2009}), where $\lambda_{i}$ is the $i^{th}$ eigenvalue of the dynamical matrix $D_{\alpha\beta} = \partial^{2} E / \partial \bm{r}_{\alpha}\partial \bm{r}_{\beta}$ and $m$ is particle mass.  Frequency $\omega_{i}$ is presented in units of $V_{p}^{1/3}\sqrt{Y/\rho}$, where $\rho$ is mass density of the solid phase.  Coordinates $\bm{r}_{\alpha}$ of particle $\alpha$ are composed of translational $x_{i,\alpha}$ and rotational $\theta_{i,\alpha}$ components $\bm{r}_{\alpha}= \{x_{1,\alpha},x_{2,\alpha},x_{3,\alpha},R_{\alpha}\theta_{1,\alpha},R_{\alpha}\theta_{2,\alpha},R_{\alpha}\theta_{3,\alpha}\}$, where $R_{\alpha}$ is the radius of gyration.  We calculate $D_{\alpha\beta}$ through central differences of forces and moments, considering only its symmetric part.  The resulting spectra of static equilibrium systems with density nearest $\phi_{J}$ are displayed in Fig.\ 1(a).  For static equilibrium systems at all densities we find $6N-3$ stable modes with $3$ trivial (rigid body) translational modes, confirming that our soft packings are indeed stably jammed as in \cite{HernPRE2003,MailPRL2009}.  This result also confirms that all $N$ particles participate in the mechanical network, and therefore no rattlers exist.  

The mean values of the $50$ lowest and highest non-trivial frequency modes $\langle \omega_{l} \rangle$ and $\langle \omega_{h} \rangle$ were computed and plotted against excess density, $\Delta \phi = \phi-\phi_{J}$ [Fig.\ 1(b)].  The scaling of $\langle \omega_{l} \rangle$ with respect to $\Delta \phi$, $\langle \omega_{l} \rangle \sim (\Delta \phi)^{2}$, reveals that these packings are in fact marginally stable at $\phi_{J}$, while finite $\langle \omega_{h} \rangle$ persists at the jamming point [Fig.\ 1(b)].  The latter behavior is a signature of translational vibrational modes involving face-face contacts, because only such contact topologies exhibit harmonic (\textit{i.e.}, quadratic) energy variation and consequently constant stiffness with respect to displacements along the direction of contact force induced by straining from $\phi_J$.  By classifying contact topologies with methods outlined subsequently we have confirmed that these modes are localized on clusters formed by face-face contacts.  Energies of these systems scale as $E \propto (\Delta \phi)^{6}$ \cite{SmiAlaPRE2010} with bulk modulus $K$ scaling as $K \propto (\Delta \phi)^{4}$.  Therefore it is clear that low frequency modes are excited by volumetric strain because $\langle \omega_{l} \rangle \propto \sqrt{K} \propto (\Delta \phi)^{2}$.


\textit{Angular alignment.} Soft contacts between faceted particles can be classified as face-face, edge-face, vertex-face, or edge-edge.  Hereafter vertex-face and edge-edge contacts are referred to as `lower order' since they exhibit less order relative to face-face and edge-face contacts.  Accordingly we define angular order metrics that approach zero as contacts orient with perfect face-face or edge-face alignment; these metrics could be measured experimentally with tomographic reconstruction.  The face-face alignment angle $\theta_{f-f,ql}$ of contacting particles $q$ and $l$ is expressed as $\theta_{f-f,ql} = \pi - \cos^{-1} ( \min ( \hat{n}_{qi}^{T}\hat{n}_{lj} ) )$, where $\hat{n}_{qi}$ is the normal vector of face $i$ on particle $q$. The minimum is computed over all combinations of $i$ and $j$ corresponding to intersecting faces on the respective particles.  The edge-face alignment angle $\theta_{e-f}$ is calculated as the minimum of $\theta_{e-f,ql}$ and $\theta_{e-f,lq}$, where $\theta_{e-f,ql} = \sin^{-1} ( \min ( |\hat{e}_{qi}^{T}\hat{n}_{lj}| ) )$, and $\hat{e}_{qi}$ is the unit edge vector of edge $i$ on particle $q$. 

Randomly oriented faces and edges provide a starting point for understanding alignment angle distributions in jammed systems.  Three-dimensional (3D) random edges and faces possess probability density of edge-face alignment with $p(\theta_{e-f}) \propto \cos (\theta_{e-f})$.  Edge-face contacts are therefore expected to be ubiquitous in jammed systems, because probability is weighted toward small $\theta_{e-f}$.  In contrast, face-face contacts are expected to be less common, because 3D random faces possess alignment probability density that vanishes in the face-face limit [$p(\theta_{f-f}) \propto \sin (\theta_{f-f})$].  On the other hand, $p(\theta_{f-f})$ for edge-face constrained contacts exhibit uniform non-vanishing probability density.

\begin{table}[ht] 
\caption{Critical edge-face $\theta_{e-f,c,i}$ and face-face $\theta_{f-f,c,i}$ alignment angles for each topology $i$ in degrees.}
\scriptsize
\begin{tabular}{cccccc}
\hline
\hline
  & \multicolumn{2}{c}{$\theta_{e-f,c,i}$} & \multicolumn{3}{c}{$\theta_{f-f,c,i}$} \\
\hline
 topology, $i$ & vertex-face & edge-edge & vertex-face & edge-edge & edge-face \\
\hline
  Tetrahedra & 54.7 & 54.7 & 70.5 & 70.5 & 54.7 \\
  Octahedra & 45.0 & 35.3 & 54.7 & 48.2 & 35.3  \\
  Dodecahedra & 20.9 & 31.7 & 37.4 & 43.6 & 31.7 \\
  Icosahedra & 31.7 & 20.9 & 37.4 & 29.2 & 20.9 \\
\hline
\hline
\end{tabular}
\end{table}

\begin{figure}
\includegraphics[width=4.1cm]{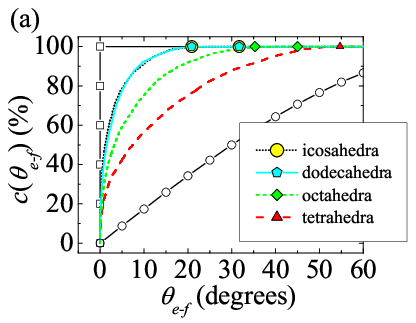}
\includegraphics[width=4.1cm]{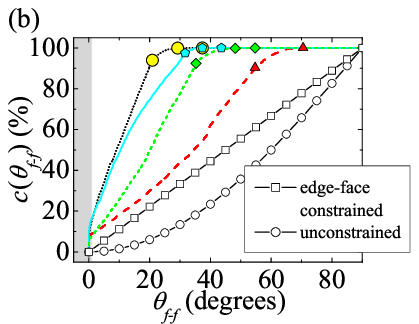}
\caption{(Color online) CDFs of (a) $\theta_{e-f}$ and (b) $\theta_{f-f}$ for stable systems nearest $\phi_J$.  Points plotted on the simulated curves correspond to $c(\theta_{e-f,c,i})$ and $c(\theta_{f-f,c,i})$.  Region 1 of $c(\theta_{f-f})$ is masked in gray, while region 2 extends from the edge of region 1 to $\theta_{f-f,c,min}$, the minimum critical face-face alignment angle.  Ideal random CDFs for edge-face constrained (open squares) and unconstrained (open circles) contacts are displayed as well.}
\end{figure}

Cumulative distribution functions (CDFs) are plotted in Fig.\ 2 for all the contacts in one realization of each system in addition to CDFs of ideal random systems.  Simulated CDFs exhibit critical angles plotted in Fig.\ 2 and listed in Table II that bound the possible alignment angles for face-vertex, edge-edge, and edge-face contact topologies; we denote critical edge-face and face-face alignment angles as $\theta_{e-f,c,i}$ and $\theta_{f-f,c,i}$, respectively, where $i$ denotes the particular contact topology.  These angles correspond to pairs of particles contacting with a particular topology oriented with highest possible symmetry.  The angular breadth of $c(\theta_{e-f})$ spans proportional to $\theta_{e-f,c,min}$, the minimum critical edge-face alignment angle among vertex-face and edge-edge contacts.  This results in particles with large $\theta_{e-f,c,min}$ having low probability density (\textit{i.e.}, CDF slope) of near edge-face contacts.  All systems exhibit hyper-random probability density at small $\theta_{e-f}$ with CDF approaching that of edge-face constrained contacts [open squares, Fig.\ 2(a)].

Fig.\ 2(b) shows that $c(\theta_{f-f})$ for all systems possess three regions of descending probability density (\textit{i.e.}, CDF slope): (1) ultra-low angle ($<1^{\circ}$) face-face contact region, (2) intermediate uniform probability region, and (3) high angle region.  Region 2 contains the largest portion of total contacts and exhibits $c(\theta_{f-f})$ similar to that of random edge-face constrained contacts [open squares, Fig.\ 2(b)], rather than unconstrained contacts [open circles, Fig.\ 2(b)].  This results from the abundance of contacts with small $\theta_{e-f}$ that exhibit CDFs approaching that of edge-face constrained contacts [open squares, Fig.\ 2(a)], rather than unconstrained contacts [open circles, Fig.\ 2(a)].  Region 2 possesses approximately invariant $p(\theta_{f-f})$ resulting in increased probability for contacts to form with acute $\theta_{f-f}$ [Fig.\ 2(b)] relative to completely unconstrained random systems [open circles, Fig.\ 2(b)].  The emergence of region 1 [gray mask, Fig.\ 2(b)] can be understood by considering the constrained rotation of an ideal edge-face contact.  During consolidation such a contact will rotate randomly about its edge through a range of $\theta_{f-f}$.  Repulsion between opposing faces at $\theta_{f-f}=0$ will prevent the contact from rotating further.  We observe the evolution of $c(\theta_{e-f})$ and $c(\theta_{f-f})$ as $\phi \rightarrow \phi_{J}$ qualitatively consistent with this idealized picture -- region 1 of $c(\theta_{f-f})$ emerges only after $c(\theta_{e-f})$ becomes hyper-random.


\textit{Constraint isostaticity and topological distributions.} To classify topologies we fit $c(\theta_{e-f})$ and $c(\theta_{f-f})$ independently with a piecewise continuous distribution function basis, $c_{f}(\theta) = H(\theta-\alpha) [a_0 - \sum\limits_{n=1}^3 a_{n} (\theta-\alpha)^{n}]$, where $a_{n}$ are fitted coefficients, $H$ is the Heaviside step function, and $\alpha$ is fitted CDF discontinuity.  All empirical CDFs have been fitted by minimizing $\int^{\theta_{i,c,min}/2}_{0} \! [c_f(\theta_i)-c(\theta_i)]^2 \ d\theta_i $, where $i$ corresponds to the particular alignment angle.  With the fitted parameter $a_0$ angular alignment cutoffs for topological classification $\theta_{cut}$ are determined such that $c(\theta_{cut})=a_0$.   Edge-face contacts are classified as those with $\theta_{e-f}<\theta_{e-f,cut}$ and $\theta_{f-f} \geq \theta_{f-f,cut}$ and face-face contacts as those with $\theta_{f-f}<\theta_{f-f,cut}$.

\begin{figure}
\includegraphics[width=4.1cm]{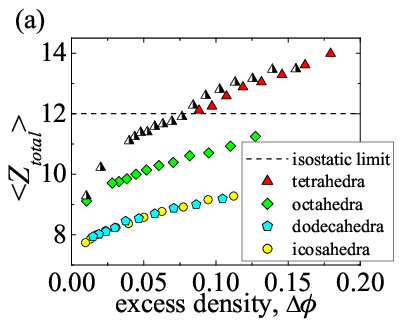}
\includegraphics[width=4.1cm]{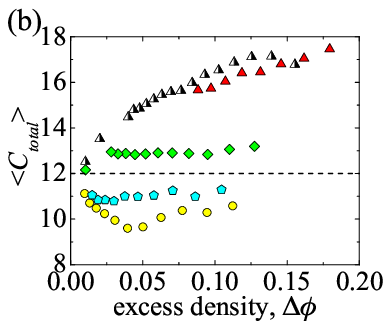}\\
\includegraphics[width=4.1cm]{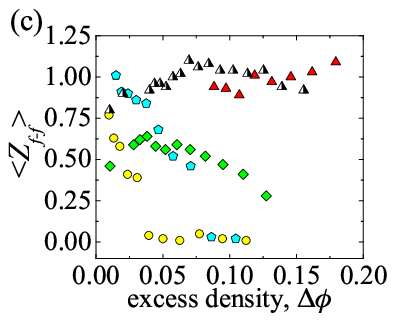}
\includegraphics[width=4.1cm]{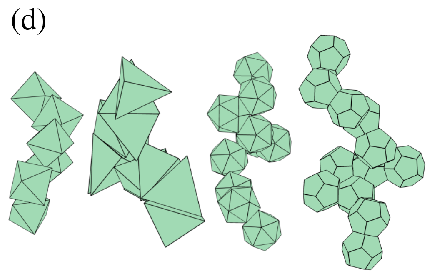}
\caption{(Color online) Variation of average (a) contact, (b) constraint, and (c) face-face contact number with $\Delta \phi$.  Half-filled symbols represent data for systems of $N=100$, while the remainder are for $N=400$.  Data points of tetrahedra and octahedra at $\Delta \phi = 0.01$  and tetrahedra at $\Delta \phi=0.02$ exhibited mild vibrational instability (5.3 \% unstable modes for tetrahedra at $\Delta \phi = 0.01$ and less than 1.2 \% for the others). (d) Face-face clusters with maximal linear extent formed in stable $N=400$ systems nearest $\phi_{J}$ of octahedra, tetrahedra, icosahedra, and dodecahedra (from left to right).}
\end{figure}

The variation of average contact number $\langle Z_{total} \rangle$ with respect to $\Delta \phi$ [Fig.\ 3(a)] confirms the generally hypostatic nature of $\langle Z_{total} \rangle$, consistent with our previous findings for smaller systems \cite{SmiAlaPRE2010}.  We determine the average contact number for face-face $\langle Z_{f-f} \rangle$, edge-face $\langle Z_{e-f} \rangle$, and lower order contacts $\langle Z_{l} \rangle$. The average constraint number $\langle C_{total} \rangle$ is thereby calculated as $\langle C_{total} \rangle = 3 \langle Z_{f-f} \rangle + 2 \langle Z_{e-f} \rangle + \langle Z_{l} \rangle$ \cite{JaoPRL2010}.  Importantly, $\langle C_{total} \rangle$ of each system approaches values near the isostatic limit [Fig.\ 3(b)], in contrast with $\langle Z_{total} \rangle$ [Fig.\ 3(a)].  Thus, these systems possess contacts that independently constrain motion.  These systems also possess $\langle Z_{f-f} \rangle \lesssim 1$ [Fig.\ 3(c)].  We attribute this effect to a two-fold rotational constraint induced on a given particle once a single face-face contact is formed.  Such a rotational constraint appears to hinder the formation of additional face-face contacts.  

\begin{table}[ht]
\caption{Distributions of average contact and constraint numbers for the present work and prior work.}
\scriptsize
\begin{tabular}{cccc}
\hline
\hline
 & present work \footnotemark[1] & various $\theta_{cut}$\footnotemark[1]$^,$\footnotemark[2] & exp. \cite{JaoPRL2010}\footnotemark[3] \\
\hline
  $\langle Z_{total} \rangle$ & 11.1 & 11.1 & 6.3\\
  $\langle Z_{f-f} \rangle$ & 0.9 & 5.4 & 2.3\\
  $\langle Z_{e-f} \rangle$ & 1.5 & 3.7 & 1.2\\
  $\langle Z_{l} \rangle$ & 8.6 & 2.0 & 2.8\\
  $\langle C_{total} \rangle$ & 14.5 & 25.5 & 12.1\\        
\hline
\hline
\end{tabular}
\footnotetext[1]{100 tetrahedra at $\Delta\phi = 0.04$.}
\footnotetext[2]{Values are computed by averaging over results obtained with $\theta_{cut}= \{5, 15, 25, 35, 45, 55\}$ as in  \cite{JaoPRL2010,JaoThesis2010}.}
\footnotetext[3]{Values presented in \cite{JaoPRL2010} for tetrahedral dice.}
\end{table}

Note that the low $\langle Z_{f-f} \rangle$ of tetrahedra clearly contrasts with the value of 2.3 recently reported for tetrahedral dice \cite{JaoPRL2010}.  The practical importance of this finding is very significant, for if two to three face-face contacts per particle were present, as reported by \cite{JaoPRL2010}, we expect such systems to readily exhibit cluster percolation and radically different structures and mechanical behavior than those with near unity $\langle Z_{f-f} \rangle$.  In Table III we compare the contact and constraint numbers of the present work (column 1) with that of \cite{JaoPRL2010} (column 3).  Indeed, all contact numbers are markedly different (except $\langle Z_{e-f} \rangle$) from that of \cite{JaoPRL2010}.  We also utilize a topological classification procedure similar to that in \cite{JaoPRL2010,JaoThesis2010} by averaging contact numbers obtained for various values of $\theta_{cut}$ (column 2 in Table III).  The resulting constraint number is twice as large as the isostatic value.  Thus, arbitrary choice of $\theta_{cut}$, as employed in \cite{JaoPRL2010,JaoThesis2010}, can yield a geometrically infeasible range of contact numbers and generally non-isostatic $\langle C_{total} \rangle$.   

\begin{table}[ht] 
\caption{Face-face contact number $\langle Z_{f-f} \rangle$, average cluster size $S$, maximal extent along Cartesian axes $l_{max}$, and the fractal dimension of clusters with maximal linear extent $D_{max}$ for stable $N=400$ systems nearest $\phi_{J}$.}
\scriptsize
\begin{tabular}{ccccc}
\hline
\hline
  & Octahedra & Icosahedra & Tetrahedra & Dodecahedra \\
\hline
  $\langle Z_{f-f} \rangle$ & 0.59 & 0.77 & 0.94 & 1.02\\
  $S$ & 1.88 & 2.45 & 3.18 & 4.13\\
  $l_{max}/(V_{cell})^{1/3}$ & 0.665 & 0.795 & 0.741 & 0.776\\
  $D_{max}$ & 1.01$\pm$0.10 & 1.23$\pm$0.08 & 1.37$\pm$0.09 & 1.32$\pm$0.14\\ 
\hline
\hline
\end{tabular}
\end{table}

We have analyzed the structure of clusters formed by face-face contacts (Table IV).  The topological connectivity of clusters was considered under periodic boundary conditions.  The average size of clusters $S$ is defined in terms of the number of particles $s$ in each cluster as $S=\sum s^2/\sum s$; we find that $S$ increases with $\langle Z_{f-f} \rangle$ as particle shape is changed.  We also find that all clusters in systems of $N=400$ particles do not percolate according to topological connectivity.  Such a requirement for periodic percolating networks is stricter than the requirement that a cluster spans the system boundaries \cite{MakPRE1995}.  Therefore, we have also considered the less restrictive percolation requirement that a cluster spans system boundaries.  Clusters in the respective systems exhibit maximal extent along Cartesian axes $l_{max}$ smaller than the length of the finite cubic cell.  Clusters with maximal linear extent exhibit chain-like structures [Fig.\ 3(d)] with fractal dimension $D_{max} \gtrsim 1$ (Table IV).  Considering the fractal dimension of $\sim 2.5$ for a percolating cluster at the threshold in a simple cubic lattice \cite{Stauffer}, these chains are substantially lower dimensional and appear to be far from percolation.  Thus, the lack of percolation is a direct consequence of near unity $\langle Z_{f-f} \rangle$.  These results suggest that face-face cluster formation is a bond percolation process with respect to $\langle Z_{f-f} \rangle$.  Therefore, we expect that systems of particles with shapes conducive to ordering (\textit{e.g.}, cubes) or with attractive interactions could increase $\langle Z_{f-f} \rangle$, forming larger clusters that percolate.


In summary, we have established the \textit{isostaticity of constraints} in disordered jammed systems of all Platonic solids except cubes and have linked this condition to their \textit{mechanical stability}.  Our results suggest that $\langle Z_{f-f} \rangle$ or other integral functions of $p(\theta_{f-f})$ are suitable order parameters that can be used to determine the \textit{maximally random jammed} state of faceted particle systems according to the approach described in \cite{TorqPRL2000}.  The structure and extent of face-face clusters is found to be a consequence of few face-face contacts in these systems.  Future work will focus on identifying means by which face-face contact number may increase and studying the critical behavior of such systems with ensuing face-face contact percolation.


\textit{Acknowledgments.} The authors acknowledge the Indo-U.S. Science and Technology Forum for supporting Purdue-JNCASR exchanges through the Joint Networked Centre on Nanomaterials for Energy (Award No. 115-2008/2009-10) and the U.S.  National Science Foundation for workshop support (Grant No. OISE-0808979) and graduate research support to K.C.S.  K.C.S. also thanks the Purdue Graduate School for financial support.  We also thank Bruce Craig and Wen Wei Loh of the Statistical Consulting Service at Purdue for helpful discussions.


\bibliography{isostatic_origins} 

\providecommand{\noopsort}[1]{}\providecommand{\singleletter}[1]{#1}%
\begin{thebibliography}{13}
\expandafter\ifx\csname natexlab\endcsname\relax\def\natexlab#1{#1}\fi
\expandafter\ifx\csname bibnamefont\endcsname\relax
  \def\bibnamefont#1{#1}\fi
\expandafter\ifx\csname bibfnamefont\endcsname\relax
  \def\bibfnamefont#1{#1}\fi
\expandafter\ifx\csname citenamefont\endcsname\relax
  \def\citenamefont#1{#1}\fi
\expandafter\ifx\csname url\endcsname\relax
  \def\url#1{\texttt{#1}}\fi
\expandafter\ifx\csname urlprefix\endcsname\relax\def\urlprefix{URL }\fi
\providecommand{\bibinfo}[2]{#2}
\providecommand{\eprint}[2][]{\url{#2}}

\bibitem[{\citenamefont{Maxwell}(1864)}]{MaxPhiMag1864}
\bibinfo{author}{\bibfnamefont{J.~C.} \bibnamefont{Maxwell}},
  \bibinfo{journal}{Phil. Mag.} \textbf{\bibinfo{volume}{27}},
  \bibinfo{pages}{294} (\bibinfo{year}{1864}).

\bibitem[{\citenamefont{Moukarzel}(1998)}]{MoukPRL1998}
\bibinfo{author}{\bibfnamefont{C.~F.} \bibnamefont{Moukarzel}},
  \bibinfo{journal}{Phys. Rev. Lett.} \textbf{\bibinfo{volume}{81}},
  \bibinfo{pages}{1634} (\bibinfo{year}{1998}).

\bibitem[{\citenamefont{Mailman et~al.}(2009)\citenamefont{Mailman, Schreck,
  O'Hern, and Chakraborty}}]{MailPRL2009}
\bibinfo{author}{\bibfnamefont{M.}~\bibnamefont{Mailman}},
  \bibinfo{author}{\bibfnamefont{C.~F.} \bibnamefont{Schreck}},
  \bibinfo{author}{\bibfnamefont{C.~S.} \bibnamefont{O'Hern}},
  \bibnamefont{and}
  \bibinfo{author}{\bibfnamefont{B.}~\bibnamefont{Chakraborty}},
  \bibinfo{journal}{Phys. Rev. Lett.} \textbf{\bibinfo{volume}{102}},
  \bibinfo{pages}{255501} (\bibinfo{year}{2009}).

\bibitem[{\citenamefont{Donev et~al.}(2007)\citenamefont{Donev, Connelly,
  Stillinger, and Torquato}}]{DonPRE2007}
\bibinfo{author}{\bibfnamefont{A.}~\bibnamefont{Donev}},
  \bibinfo{author}{\bibfnamefont{R.}~\bibnamefont{Connelly}},
  \bibinfo{author}{\bibfnamefont{F.~H.} \bibnamefont{Stillinger}},
  \bibnamefont{and} \bibinfo{author}{\bibfnamefont{S.}~\bibnamefont{Torquato}},
  \bibinfo{journal}{Phys. Rev. E} \textbf{\bibinfo{volume}{75}},
  \bibinfo{pages}{051304} (\bibinfo{year}{2007}).

\bibitem[{\citenamefont{Jaoshvili et~al.}(2010)\citenamefont{Jaoshvili, Esakia,
  Porrati, and Chaikin}}]{JaoPRL2010}
\bibinfo{author}{\bibfnamefont{A.}~\bibnamefont{Jaoshvili}},
  \bibinfo{author}{\bibfnamefont{A.}~\bibnamefont{Esakia}},
  \bibinfo{author}{\bibfnamefont{M.}~\bibnamefont{Porrati}}, \bibnamefont{and}
  \bibinfo{author}{\bibfnamefont{P.~M.} \bibnamefont{Chaikin}},
  \bibinfo{journal}{Phys. Rev. Lett.} \textbf{\bibinfo{volume}{104}},
  \bibinfo{pages}{185501} (\bibinfo{year}{2010}).

\bibitem[{\citenamefont{Smith et~al.}(2010)\citenamefont{Smith, Alam, and
  Fisher}}]{SmiAlaPRE2010}
\bibinfo{author}{\bibfnamefont{K.~C.} \bibnamefont{Smith}},
  \bibinfo{author}{\bibfnamefont{M.}~\bibnamefont{Alam}}, \bibnamefont{and}
  \bibinfo{author}{\bibfnamefont{T.~S.} \bibnamefont{Fisher}},
  \bibinfo{journal}{Phys. Rev. E} \textbf{\bibinfo{volume}{82}},
  \bibinfo{pages}{051304} (\bibinfo{year}{2010}).

\bibitem[{\citenamefont{O'Hern et~al.}(2003)\citenamefont{O'Hern, Silbert, Liu,
  and Nagel}}]{HernPRE2003}
\bibinfo{author}{\bibfnamefont{C.~S.} \bibnamefont{O'Hern}},
  \bibinfo{author}{\bibfnamefont{L.~E.} \bibnamefont{Silbert}},
  \bibinfo{author}{\bibfnamefont{A.~J.} \bibnamefont{Liu}}, \bibnamefont{and}
  \bibinfo{author}{\bibfnamefont{S.~R.} \bibnamefont{Nagel}},
  \bibinfo{journal}{Phys. Rev. E} \textbf{\bibinfo{volume}{68}},
  \bibinfo{pages}{011306} (\bibinfo{year}{2003}).

\bibitem[{\citenamefont{Gogu}(2005)}]{GogMMT2005}
\bibinfo{author}{\bibfnamefont{G.}~\bibnamefont{Gogu}}, \bibinfo{journal}{Mech.
  Mach. Theory} \textbf{\bibinfo{volume}{40}}, \bibinfo{pages}{1068 }
  (\bibinfo{year}{2005}).

\bibitem[{\citenamefont{Baker and Kudrolli}(2010)}]{BakKudPRE2010}
\bibinfo{author}{\bibfnamefont{J.}~\bibnamefont{Baker}} \bibnamefont{and}
  \bibinfo{author}{\bibfnamefont{A.}~\bibnamefont{Kudrolli}},
  \bibinfo{journal}{Phys. Rev. E} \textbf{\bibinfo{volume}{82}},
  \bibinfo{pages}{061304} (\bibinfo{year}{2010}).

\bibitem[{\citenamefont{Jaoshvili}(2010)}]{JaoThesis2010}
\bibinfo{author}{\bibfnamefont{A.}~\bibnamefont{Jaoshvili}},
  \bibinfo{type}{{Ph.D.} thesis}, \bibinfo{school}{New York University}
  (\bibinfo{year}{2010}).

\bibitem[{\citenamefont{Watanabe}(1995)}]{MakPRE1995}
\bibinfo{author}{\bibfnamefont{M.~S.} \bibnamefont{Watanabe}},
  \bibinfo{journal}{Phys. Rev. E} \textbf{\bibinfo{volume}{51}},
  \bibinfo{pages}{3945} (\bibinfo{year}{1995}).

\bibitem[{\citenamefont{Stauffer}(1985)}]{Stauffer}
\bibinfo{author}{\bibfnamefont{D.}~\bibnamefont{Stauffer}},
  \emph{\bibinfo{title}{Introduction to Percolation Theory}}
  (\bibinfo{publisher}{Taylor and Francis}, \bibinfo{year}{1985}).

\bibitem[{\citenamefont{Torquato et~al.}(2000)\citenamefont{Torquato, Truskett,
  and Debenedetti}}]{TorqPRL2000}
\bibinfo{author}{\bibfnamefont{S.}~\bibnamefont{Torquato}},
  \bibinfo{author}{\bibfnamefont{T.~M.} \bibnamefont{Truskett}},
  \bibnamefont{and} \bibinfo{author}{\bibfnamefont{P.~G.}
  \bibnamefont{Debenedetti}}, \bibinfo{journal}{Phys. Rev. Lett.}
  \textbf{\bibinfo{volume}{84}}, \bibinfo{pages}{2064} (\bibinfo{year}{2000}).

\end{thebibliography}

\end{document}